\documentclass[11pt,A4paper]{article}
\usepackage[top=0.85in,left=1.25in,footskip=0.75in,marginparwidth=2in]{geometry}

\usepackage[utf8]{inputenc}

\raggedright
\usepackage{array}

\makeatletter
\renewcommand{\@biblabel}[1]{\quad#1.}
\makeatother

\usepackage{graphicx}
\begin{document}
\title{A proposed method for a photon-counting laser coherence detection system to complement optical SETI}
\author{David M.Benton }


\maketitle
\begin{center}
Aston Institute of Photonic Technologies, Aston University, Aston Triangle, Birmingham, UK. B4 7ET.
d.benton@aston.ac.uk
\end{center}
\begin{abstract}
The detection of laser radiation originating from space is a positive indicator of Extra Terrestrial Intelligence (ETI). Thus far the optical search for ETI (OSETI) has looked for enhanced brightness in the form of either narrow-band spectral emission or time correlated photons from laser pulses. In this paper it is proposed to look for coherence properties of incoming light in a manner that can distinguish against atomic emission lines. The use of photon sensitive detectors and a modulating asymmetric interferometer has been modelled. The results suggest that the sensitivity of detection for continuous laser sources would be better than current spectroscopic approaches, with detection thresholds of $<$1 photon s$^{-1}$m$^{-2}$ against a background with a spectral bandwidth of 0.1nm over an observation time of 750s.
\end{abstract}
photon detection, coherence, SETI

\section{Introduction}
Laser radiation does not occur naturally, thus any detection of laser radiation implies an intent to put this radiation to use. In astrophysical terms the detection of laser radiation is a strong indicator of intelligence and technology - a so called technosignature \cite{tarter2006evolution}. This was pointed out at the early inception of the maser \cite{SCHWARTZ1961}. There are many potential technosignatures that have been actively searched for  \cite{TarterReview} but the predominant medium of searches for extra terrestrial intelligence (ETI) signals has been the radio spectrum. Recent years have seen more effort directed towards optical detection methods \cite{Howard2007} \cite{Werthimer2001} \cite{Reines2002} \cite{Tellis2017}. A particular advantage arises from the low divergence of higher frequency beams - such as optical and infra red lasers – which concentrates the signal making it easier to detect but only if it is well directed \emph{i.e.} there is an intent. It also reduces the required power in the laser making it technologically easier to implement. Detecting any laser signal has relied on the enhanced spectral brightness of the laser or the enhanced temporal brightness that pulsed lasers achieve. 
Detection of  continuous lasers by their spectroscopic signature involves looking for peaks of enhanced brightness against blackbody emission containing absorption dips  \cite{Reines2002},\cite{Tellis2017},\cite{stone2005lick} which has involved algorithmic based searches of stellar spectra in collected databases.  They concentrate on looking for narrow emission peaks amidst high resolution absorption spectra. Given the spectral resolution and the detector sensitivity it is estimated that ETI lasers in the MegaWatt range would be detected with powers of 10's of kW also possibly detectable in some circumstances.  As yet no laser signatures have been found. 

Arguably the preferred approach has been to look for short optical pulses - modulation being another technosignature - which would be created by a laser to achieve the required brightness more easily than with continuous sources.  Short pulses (10nS or less) can be observed by looking for two and three way coincidences in photon sensitive detectors  \cite{Werthimer2001}  \cite{Maire2014} \cite{schuetz2016optical}, \cite{schuetz2018recent}. This has been performed over a number of years with telescopes typically of a scale 1m  \cite{wright2001improved},  and would expect to see photon densities of around 10 photons per pulse. Abeysekara \cite{abeysekara2016search} made alternate use of a larger ground based gamma observatory to detect optical pulses with a sensitivity of 1 photon $m^{-2}$. 
\begin{table}[]
    \centering
    \begin{tabular}{| m{3cm} m{2cm} m{2cm}  m{1.7cm}  m{1.5cm}  m{2cm}|}
    \hline
         Observation &   Collecting Area $m^2$ & Exposure time & Sensitivity (phot $m^-2$)
         & Laser type \\ 
         \hline
        Tellis and Marcy \cite{Tellis2017} & 76 & 600s & 1 & P+CW \\ 
        \hline
         Reines and Marcy \cite{Reines2002}& 76 & &100 & P+CW  \\
         \hline
       Wright \cite{wright2001improved}  & 0.78 &5ns &51 & P  \\
       \hline
        Howard \cite{howard2004search}& 1.7 & 5ns &100 &P  \\
        \hline
         Hanna \cite{hanna2009oseti}&2300 & 12ns &10 &P  \\
         \hline
         Abeysekara \cite{abeysekara2016search} & 110 & 12ns & 0.95& P  \\
         \hline
         Schuetz \cite{schuetz2016optical} & 0.2 & 25ns & 67 & P  \\
         \hline
    \end{tabular}
    \caption{Indicative performance characteristics from Optical SETI searches. P represents pulsed laser detection, CW is continuous wave detection.}
    \label{tab:Sensitivity}
\end{table}
 
Borra \cite{Borra2012} considered that the effect of a train of short optical pulses upon a detected spectrum would be to produce an optical comb in frequency space. This would show up as a periodic signal when the spectrum was Fourier transformed. Indeed such signals were observed \cite{Borra2016} but their origin is yet to be established with Hippke \cite{Hippke_2019} postulating that they arise from non-random spectral absorption features.   
Leeb \cite{Leeb2013} considered detecting periodic pulsed signals in the temporal domain, but ruled out looking for continuous signals: 
\emph{"We rule out a continuous signal (i.e., zero modulation), as this cannot easily be discriminated against natural light sources".} The case for using coherence when using heterodyne or homodyne detection techniques was delivered robustly by Kingsley \cite{Kingsley1993}. 

 The detection techniques used are looking for different things - pulse detection cannot detect continuous sources. Spectroscopic detection integrates over time and cannot distinguish pulsed from continuous lasers. A table of indicative performance values is  given in table \ref{tab:Sensitivity}, with more example given in  \cite{schuetz2016optical}. The two techniques are not directly comparable as the pulsed detection technique can in principle detect a single pulse which would be significant when the false alarm rate is low - as is the case with more than 2 detectors and the relevant quantity is the number of photons per pulse per square meter of detection area. This signal does not integrate due to the threshold of detection for coincidences. The spectroscopic technique looks for signal not against the received flux but against the photon noise and therefore requires integration. There is then a threshold of received photons per square meter per second relative to the background, which is dependent upon the star type and the spectral bandwidth.
 It is proposed in this paper that a method for detecting truly coherent, continuous laser sources is a realistic and effective way to look for ETI laser signals. This technique is most comparable to the spectroscopic observations as it requires integration but has the potential for flexibility in detecting both continuous and pulsed sources in post processing of events.
 
 \section{Method}
 
Earthly laser detection systems tend to concentrate on the intensity characteristics of laser radiation, that is to look for enhanced spectral intensity or high temporal brightness and the same is true for OSETI. Recently a technique based on detecting coherence has been used by Benton\cite{Benton2017} to positively detect faint laser radiation against a bright incoherent background, which is of course the same problem faced by OSETI. The utilisation of coherence with OSETI was mentioned by Kingsley \cite{Kingsley1993} who discussed the possibility of heterodyne detection. He also mentioned homodyne detection but dismissed it due to unspecified difficulties.  
\begin{figure}
    \centering
    \includegraphics{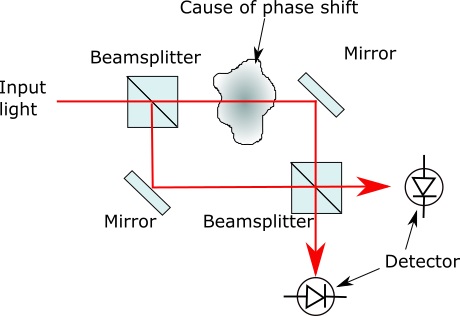}
    \caption{A schematic diagram representing a Mach Zehnder interferometer .}
    \label{fig:mzi}
\end{figure}
The concept being proposed here involves the use of an asymmetric Mach Zehnder interferometer (MZI). The MZI divides incoming wavefronts along 2 paths using a beamsplitter, which are subsequently recombined with a second beamsplitter \cite{hecht2002optics} shown schematically in figure \ref{fig:mzi}. If in travelling along either path the optical phase is shifted then this manifests itself as an intensity variation when the light from the 2 arms is recombined and then detected. The detected intensity varies according to:
\begin{equation}
    I_d(t)=\frac{I_0}{2}(1+G(\Delta\lambda)cos(\frac{2\pi}{\lambda}\phi(t)))
\end{equation}
Where $I_{d}(t)$ is the time varying detected intensity, $\lambda$ is the optical wavelength and $\phi(t)$ is the time varying phase occurring in one of the interferometer arms. $G(\Delta\lambda)$ is a complex temporal coherence function related to the spectral power density of the source \cite{goodman2015statistical}. A star with a wide spectral bandwidth has a very short coherence length ($<$1$\mu$m) whereas lasers typically have coherence lengths of many cms or more.  The approach taken by Benton for laser detection was to introduce a fixed path length difference between the interferometer arms and then to sinusoidally modulate the length of one of the arms with a piezo-mounted mirror. The modulation of the path length (a temporal change of relative phase $\phi$(t)) gives rise to an intensity modulation \emph{only} if the source coherence length exceeds the fixed path length difference.  Thus the detection of a modulating signal implies the presence of a coherent source.  

 This technique can be applied to the search for ETI lasers with account taken of the intensities and timescales involved. Looking for continuous ETI sources draws comparisons with a spectroscopic detection approach.  Tellis and Marcy \cite{Tellis2017} indicate there are issues that arise as a consequence of using an Echelle spectrograph such as found on the Keck Telescope used to look for laser emission in stellar spectra. Firstly there are examples of atomic emission lines being detected such as can happen in solar flares \cite{Singh2006}. These would be indistinguishable from laser emission as their spectral width is less than the resolution of the spectrometer. Such emissions are identified by a spectral analysis algorithm and must be discarded by manual intervention. Secondly  the long integration time sees rapid events such as cosmic ray interactions with the spectrograph CCD leaving a localised signature which can also give a false positive to the analysis algorithm. This too can be excluded after manual  examination. In principle both of these issues would not cause a problem for a suitably designed coherence detection system.   A third issue is the need for a reference spectrum to be taken in order to observe changes. Other techniques such as multi pulse coincidence detection will work only if the laser source is pulsed.  
Applying the concept of coherence detection using interferometry would involve looking for a modulation of the output intensity.  This modulation is observed by sampling the output signal and extracting an amplitude at the modulation frequency. In the case of stellar interferometry the photon flux is so low that photon sensitive detectors – such as avalanche photodiodes (APDs) are required. In this paper a new method for coherence detection with a photon sensing coherence detector using an MZI is described. At each of the 2 output ports of the MZI is located a photon sensitive detector such as a silicon APD. Every photon detected is recorded with its time of detection accurate to sub 1nS – called a time tag.  Processing can be done off line using the time tags. Path length modulation can be provided by a piezo mounted mirror in one of the arms driven with a sinusoidal signal. The amplitude of the modulation causes a phase shift which is of course wavelength dependent and so must be chosen to maximise the signal for expected signal wavelength. In order to discriminate emission lines the requirement for the asymmetry in path length difference between the arms must exceed the expected coherence length, which can be determined from the emission line width.  
The coherence length $l_c$ of an emission with a Gaussian profile is given by :
\begin{equation}
    l\small{c}=\sqrt{\frac{2ln2}{\pi}}\frac{\lambda^2}{\Delta\lambda}
\end{equation}

Where $\lambda$ is the source wavelength and $\Delta\lambda$ is the spectral linewidth \cite{akcay2002estimation}.  Examining the spectral width of the strongest Fraunhofer lines from the sun the narrowest line has a width of 0.006nm (Fe I 525.0216nm) which has a coherence length of 43 mm. This however is not an emission line and most lines are much broader than this. Emission lines can arise from coronal loops \cite{Singh2006} where the narrowest observed lines are 0.05nm. Thus coherence lengths for wavelengths around 656nm (H$\alpha$) would be around 8mm \footnote{The Doppler effect of Earth's motion does not affect the linewidth as this introduces a positional shift. This shift can manifest itself as an effective width increase in the spectroscopy methods}. On the very reasonable assumption that lasers have narrower linewidths than this then introducing a path length difference of 10mm or more should prevent emission lines from giving rise to a modulating output and only lasers would cause intensity modulation.  

The expected stellar background intensity levels can be estimated in relation to the solar photon flux, which is approximately 4x10$^{18}$ photons s$^{-1}$m$^{-2}$ nm$^{-1}$ in the visible region \cite{SMESTAD2008371}. The photon flux scales with the inverse square of the distance so for a sun-like star at a distance of 1kPc the received flux would be
\begin{equation}
    F=F_{s} (\frac{1au}{1kPc})^2 = 94 s^{-1} m^{-2} nm^{-1}   
\end{equation}
where $F_{s}$ is the solar photon flux. The total received background photons are calculated from 
\begin{equation}
   N= F A t \Delta\lambda 
\end{equation}
where A is the telescope receiver area, t is the observation time and $\Delta\lambda$ is the spectral bandwidth.  
The OSETI spectral examinations made using the 10m Keck telescope provide us with measurement of expected level of stellar background photons \cite{Tellis2017}. The Echelle spectrometer used has a pixel resolution of $\lambda/\Delta\lambda$=250,000 and is operated for around a 10min observation time to give somewhere between 10,000 to 40,000 photons per pixel (dependent upon the S/N required). Choosing an in-between value of 30,000 photons in 600s this gives around 50 photons $s^{-1}$.  At a wavelength of 500nm 1 pixel represent 0.002nm, this gives 500 pixels per nm and of the order 25,000 photons $s^{-1} nm^{-1}$. This is the same order as the estimated photon flux in equation (3) which gives 9000 photons $nm^{-1}$ but will vary significantly depending upon the stellar luminosity and distance. As is the case for spectral measurements, any laser signal needs to be discerned against the photon noise not against the absolute background signal.

The intensity at a detector placed at one of the output ports of the MZI follows a cosinusoidal modulation \cite{hecht2002optics}\cite{Benton2017} as described in equation 1. 
\begin{equation}
    I_d=I_0\gamma(\lambda)[1+cos(k(\Delta L(t)))]
\end{equation}

Where $I_0$ is the input intensity,  $\gamma(\lambda)$ is a wavelength dependent factor representing reflection coefficients at both beamsplitters and also incorporates losses, $k =2\pi/\lambda$  and $\Delta L$ is the difference between the length of the interferometer arms  $L_1-L_2(t)$. 

Modulating the arm length with a piezo mirror with a modulation voltage of amplitude $v_m$ at a frequency $f_m$  the path length difference is 
\begin{equation}
    \Delta L(t)=L_1-L_2(0)-v_mp\sqrt{2}sin(2\pi f_mt)-v_{off}p\sqrt{2}
\end{equation}
Where $p$ is the response of the piezo in $\mu m/V$and $v_{off}$ is an offset voltage. With careful choice of offset voltage the intensity at a detector can be written as 
\begin{equation}
    I_d(t) = I_0\gamma (\lambda)[1+sin(k v_mp\sqrt{2}sin(2\pi f_mt))]
\end{equation}
Sampling the detector signal over a period of time and taking the Fourier transform shows power concentrated in the modulation frequency and is an indicator of coherent input. The signals at each of the output ports are 180$^{\circ}$ out of phase thus subtracting the 2 signals increases the S/N.

\section{Results}
\begin{figure}[h!]
    \centering
    \includegraphics{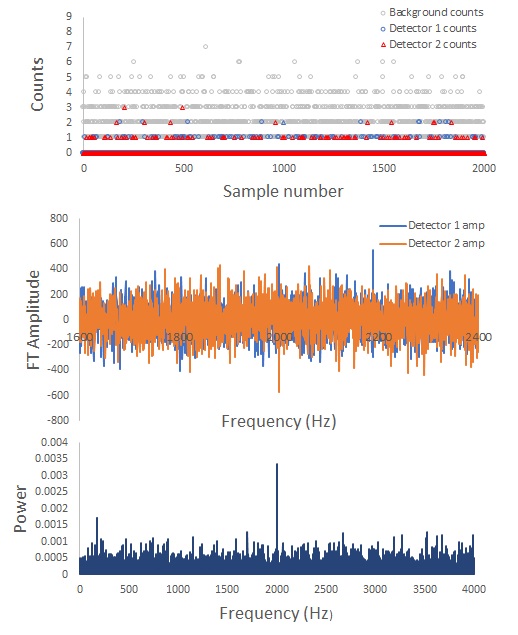}
    \caption{Examples of the generated data used. The upper plot is a subset of the count data showing the individual counts generated as background and the laser signal photons in each detector. The background rate was 25k photons $s^{-1}$ and the laser signal rate is 500 photons $s^{-1}$. The modulation frequency was 2kHz. The centre plot shows the FT amplitudes from both detectors around the modulation frequency with the amplitudes being of opposite phase. The lower plot is the power from the subtraction of the detector amplitudes clearly showing the signal at 2kHz. }
    \label{fig:FTplot}
\end{figure}

Estimation of detectability of laser signals was performed by modelling Poissonian photon count distributions for noise and laser signal. Within the model the modulation frequency of a piezo mounted mirror was chosen to be 2kHz for the following reasons. Source intensity modulations of incoherent background appear as modulations in the detector outputs and this can arise from atmospheric scintillation \cite{2015MNRAS}.
The atmospheric scintillation frequency spectrum extends to around 1kHz so the modulation frequency is placed beyond this level to reduce background noise. This is dependent on telescope size and is less prevalent in large telescopes.
With a sparse photon rate the number of events per cycle is best maximised to reduce the high frequency contributions of discrete event based signals \cite{Kingsley1993}. It was found that sampling at 8kHz worked well. The noise and signal Poisson distributions were produced from the average photon detection rate per sample period based upon an expected photon level for the Keck telescope as discussed earlier (25000/s background with 8kHz sample rate has an average of 0.31 photons per sample).
The laser signal distribution was multiplied by the interferometer modulation function $\frac{1}{2}(1+cos(2\pi f_mt))$ and then added to the noise. In the case of using 2 detectors, one at each output port of the MZI, the signal photons are directed to particular detectors depending on the phase of the modulation function. This was determined by comparing the absolute amplitude of the modulation function with a function 180$^\circ$ out of phase with it.   Dark count rates in APDs are so low (and non Poissonian) as to be negligible and were not included. There are a number of ways to analyse a set of photon events but here consistency is kept with  the author's earlier approach of summing FFTs \cite{Benton2017}. A data set of counts per sample time for each detector $x_n(t)$ was Fourier Transformed ($\mathcal{F}(x_n)$) and then these two transform sets were subtracted, and the magnitude of the result taken  
\begin{equation}
[\mathcal{F}(x_1)-\mathcal{F}(x_2)][\mathcal{F}(x_1)-\mathcal{F}(x_2)]^*
\end{equation}
Example plots of the generated counts and the FT signals are shown in figure{\ref{fig:FTplot}}. This data shows a subset of a 20,000 sample set where the background rate was 25k photons $s^{-1}$ and the laser signal rate is 500 photons $s^{-1}$ distributed between the two detectors (this is chosen to give observable signal from a single set for demonstration purposes). The sample rate was 8kHz and the modulation frequency was 2kHz. A section of the FT amplitude spectrum is given in the central plot with the modulation frequency in the centre of the region. It can be seen that the amplitudes at the modulation frequency show opposite signs in the 2 detectors because they are out of phase. The lower plot shows the resulting power spectrum from the subtraction of the amplitude spectra from the two detectors with a clear peak at the modulation frequency.The height of this peak is extracted and compared with the noise level throughout the frequency space. 

\begin{figure}
    \centering
    \includegraphics[width=\textwidth]{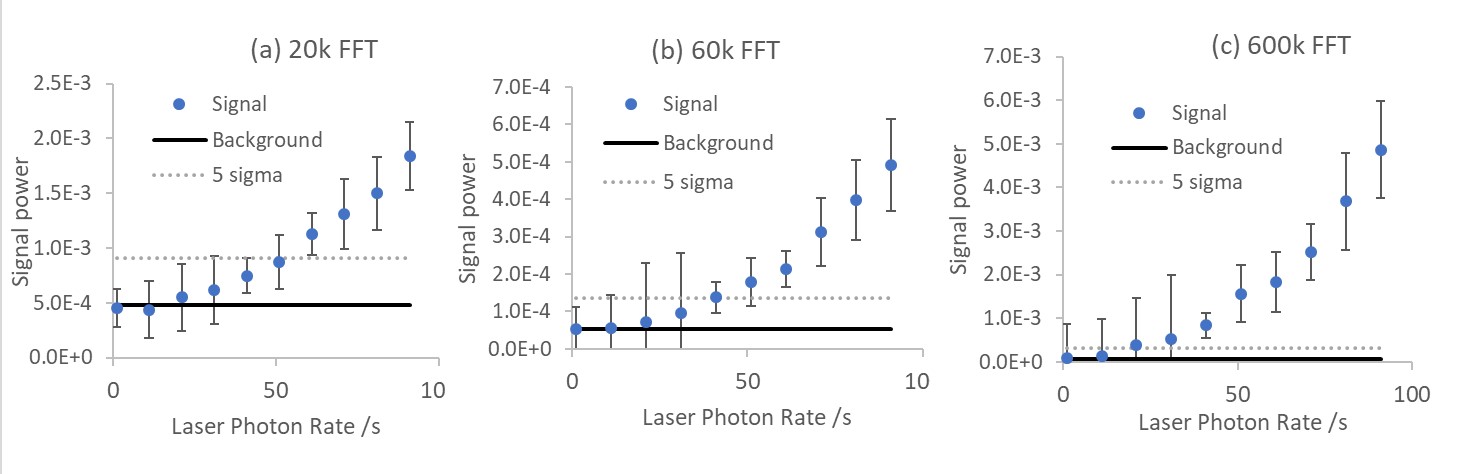}
    \caption{Signal detection level including background noise level for FFT of different sizes. In each case the cumulative observation time is 750s. The dotted line represents a statistical level of 5 sigma. }
    \label{fig:NumPoints}
\end{figure}

\begin{figure}[h]
    \centering
    \includegraphics[width=\textwidth]{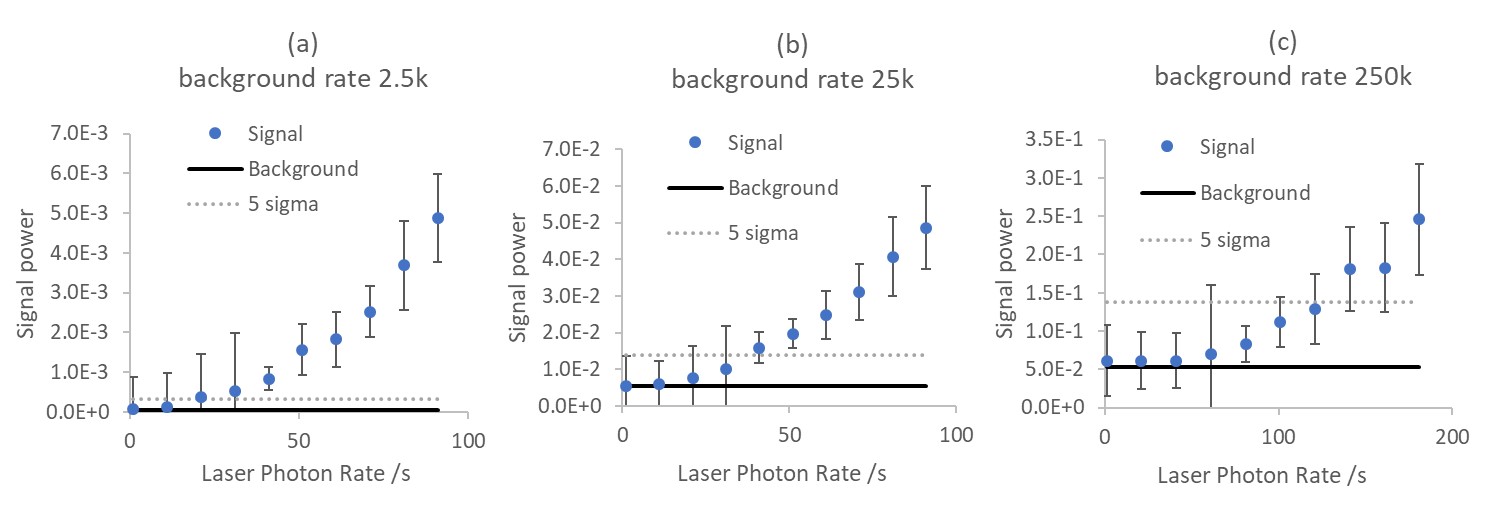}
    \caption{Signal detection level including background noise level for background count rates (a)  2.5k per second (equivalent spectral bandwidth of 0.1nm ) (b) 25k per second (1nm)  and  (c) 250k per second (10nm).Total observation time for each plot is 750s. The dotted line represents a statistical level of 5 sigma.}
    \label{fig:background}
\end{figure}

This was repeated and sets of magnitude values were summed with the power level at the modulation frequency extracted along with the background level. Initially sample sets of 20,000 points were transformed with 300 sets summed yielding an observation time of 750s, a similar scale to the spectroscopy measurements \cite{Tellis2017}. Modelling explored the effect of sample size and spectral bandwidth on the detectability of laser signals of varying amounts. 
Increasing the size of the sample set increases resolution and thereby improves S/N as long as the reference modulation is stable. This can be seen in the plots shown in figure \ref{fig:NumPoints} where the better resolution resulting from a larger number of points gives rise to stronger signals. These plots all represent an observation time of 750s. In this case the potential laser signal rate was varied and repeated 10 times to estimate the signal level uncertainty. The dotted line on the plots represents 5 standard deviations from the mean level of the background and is treated as a reliable detection threshold. The background rate was set at 2.5k to represent a 0.1nm spectral bandwidth in a Keck sized telescope. A 600k point FFT results in an improved detection threshold cf an equivalent observation time with 20k FFT sets.  A single FFT for the entire data set  - 6 million sample points would represent the ultimate resolution and be equivalent to the sensitivity obtained from a lock in amplifier with a long integration time.
The spectral bandwidth determines background photon rate and also the photon noise against which any laser signal must be detected. A spectral bandwidth of 1nm would be roughly equivalent to 25k counts per second. Models of the signal level vs the incident laser photon rate were conducted for 0.1nm, 1nm and 10nm spectral bandwidths (assuming perfect collection and detection). This can be seen in figure \ref{fig:background} where the detectable laser signal level is around 130 photons s$^{-1}$ with a 10nm bandwidth, 40s$^{-1}$ with a 1nm bandwidth and around 20s$^{-1}$ with a 0.1nm bandwidth. These plots were made using 600k point FFTs summed to give a total observation time of 750s. 

\begin{figure}[h!]
    \centering
    \includegraphics[width=\textwidth]{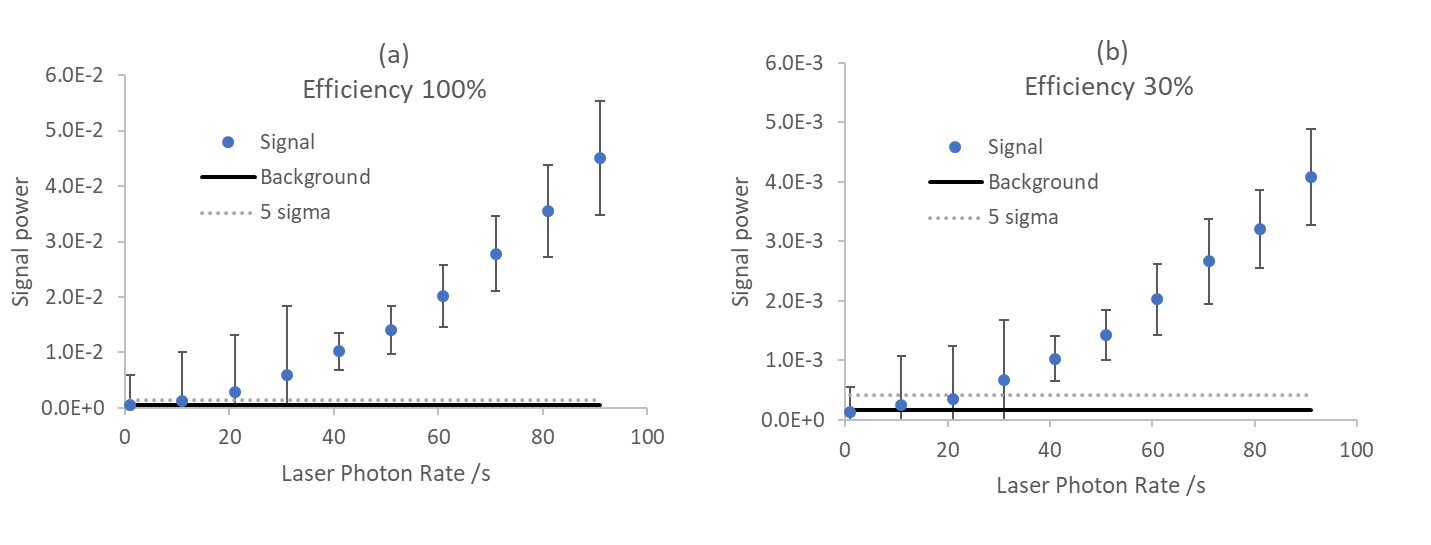}
    \caption{Signal detection level including background noise level for 600k point FFT with the assumption of (a) perfect efficiency and (b) 30\% efficiency. The dotted line represents a statistical level of 5 sigma.  }
    \label{fig:EfficiencyPlot}
\end{figure}

This modelling so far has assumed perfect optics and detectors. The effect of loss and detector quantum efficiency is shown in the plots of figure \ref{fig:EfficiencyPlot} where a comparison is made between perfect collection and 30\% efficiency.  
Some of the most sensitive scenarios seen here have a sum of 10 sets of 600k point FFTs (750s observation time) showing a detectable signal with 10-20 laser photons/s (100\% efficiency) with a background level of 2500$s^{-1}$ equivalent to a filtered spectral bandwidth of 0.1nm, a total of some 15,000 photons. This approach as a whole compares favourably with the sensitivity of the spectroscopy approach \cite{Tellis2017} which expects a threshold of 50,000 photons and claims 83 photons per second threshold for pulse detection with a typical 10 minute observation. Expected sensitivities are shown in table\ref{tab:my_results}.
It is worth considering that the use of a Keck type telescope represents a high level bespoke investigation, whereas there is a more accessible scale of investigation which is more broadly applicable. Modelling for a reduced scale telescope with a 1m diameter mirror, assuming the same parameters (600k FFT, 750s observation time) results in a detection threshold of around 400 photons $s^{-1}$ using a 1nm spectral bandwidth. 

\begin{table}[]
    \centering
    \begin{tabular}{|m{2cm} m{1.5cm}  m{1.8cm} m{1cm}  m{1cm} m{1cm} m{1cm} m{1cm}  m{1cm} m{1.5cm}|}
    \hline\hline
          Background rate($s^{-1}$) & Laser 5$\sigma$($s^{-1}$) & Sensitivity ($s^{-1}$/m$^{2}$) & Laser photons & B photons & Poisson noise & 5x noise & Ratio 5$\sigma$:5 Poisson \\
         \hline
2.5k  &20	&0.26	&15000	&1.87E6  &1369	&6846	& 2.19 \\\hline
25k	  &40	&0.52  &30000	&1.87E6	 &4330	&21650	& 1.38\\\hline
250k  &130	&1.71	&97500	&1.87E8	&13693	&68465	&1.42 \\\hline
2.5k	&20	&0.26	&15000	&1.87E6	&1369	&6846	&2.19\\\hline
330$^a$		&5.3 &5.3	&3947	&2.5E5	&500	&2500	&1.59\\\hline
750$^b$	 	&15	&0.20	&11250	&5.6E5	&750	&3750	&3\\\hline

    \end{tabular}
    \caption{Data showing the relative sensitivity. All data was produced for a 600k sample FFT with 10 iterations resulting in an observation time of 750s. The background rate (col 1) scales with spectral width at 25k per nm. The  Laser 5$\sigma$ column is the laser photon rate needed to give a positive signal. Sensitivity is the 5$\sigma$ rate divided by the detecting area. The final ratio column is the ratio of detected photons to the expected Poisson $5\sigma$ level.
    $^a$Represents a telescope with a collecting area of 1$m^2$.
    $^b$Represents the large Keck system with a detection efficiency of 30\%. }
    \label{tab:my_results}
\end{table}

\begin{figure}[h]
    \centering
    \includegraphics[width=\textwidth]{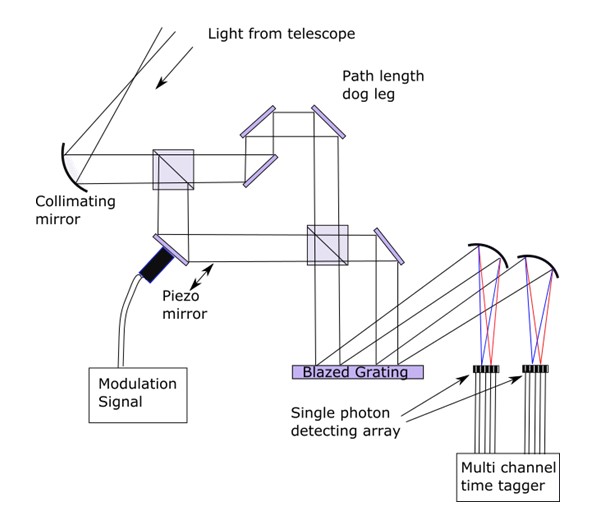}
    \caption{A schematic diagram for an asymmetric MZI with a dual output port array of APDs following spectral dispersion from a grating. }
    \label{fig:setup}
\end{figure}
A system to implement this concept could operate as follows and is shown in figure \ref{fig:setup}. A MZI with an extended length arm must sit in a section of collimated light, as could be provided by a parabolic mirror post collection telescope. The required spectral resolution can be obtained using a diffraction grating directing dispersed light onto 2 arrays of single photon detectors. The collimated light from both arms should interact with the same diffraction grating to ensure consistent resolution between detector arrays.  An array able to give complete spectral coverage would be a large and involved system with potentially thousands of independent detection and timing channels replicated at both output ports. This would represent the state of the art in APD arrays (\cite{bruschini2017ten}) but an array of 1024 APDs is certainly feasible. It is worth noting that the throughput of photon detections is the same no matter the number of channels (provided they are not saturated), but more channels means better resolution and reduced spectral bandwidth per channel which reduces the threshold for detection. Providing wide spectral coverage within a single detector is currently beyond single photon detector technology and several detection wavebands would be required with different APD technologies. A more targeted approach may be required making use of the “magic colours for OSETI” \cite{Narusawa2018},\cite{Hippke2018}. This could use small arrays of APDs with quantum efficiencies matched to the required wavelength. Such wavelengths have been suggested matching Nd:YAG laser wavelengths (a rather anthropomorphic approach). Other suggestions include using specific absorption wavelengths which would be universal but may be less intense lasers. A system which has a tunable wavelength filter could be considered but may require excessive observation time for comprehensive spectral coverage.

\section{Discussion}
\subsection{Power Requirements}

Assuming the transmit laser originates from equivalent optical systems to our own, then the optimistic beam size is determined by the diffraction limit with the diffraction angle $\theta=2.44\frac{\lambda}{D_t}$ where $\lambda$ is the wavelength and $D_t$ is the diameter of a circular transmitting aperture. The received photon flux in photons s$^{-1}$ is :
\begin{equation}
    n=G \varepsilon(\lambda)\frac{P}{E(\lambda)}
\end{equation}
where G is the geometric loss factor 
\begin{equation}
    G=\frac{D_t^2 D_r^2}{2.44 \lambda d^2}
\end{equation}
$D_r$ is the receiver diameter and d is the star to earth distance. The factor $\varepsilon$ is a wavelength dependent loss factor accounting for extinction and atmospheric absorption (0.001kPc$^{-1}$) \cite{Hippke2018}, $P$ is the transmitted laser power and $E$ is the photon energy $E=hc/\lambda$ where $h$ is Plancks constant and $c$ is the speed of light. This is a standard interpretation given in several forms \cite{Clark2018}\cite{Hippke2018}.
\begin{figure}[h]
    \centering
\includegraphics[width=\textwidth]{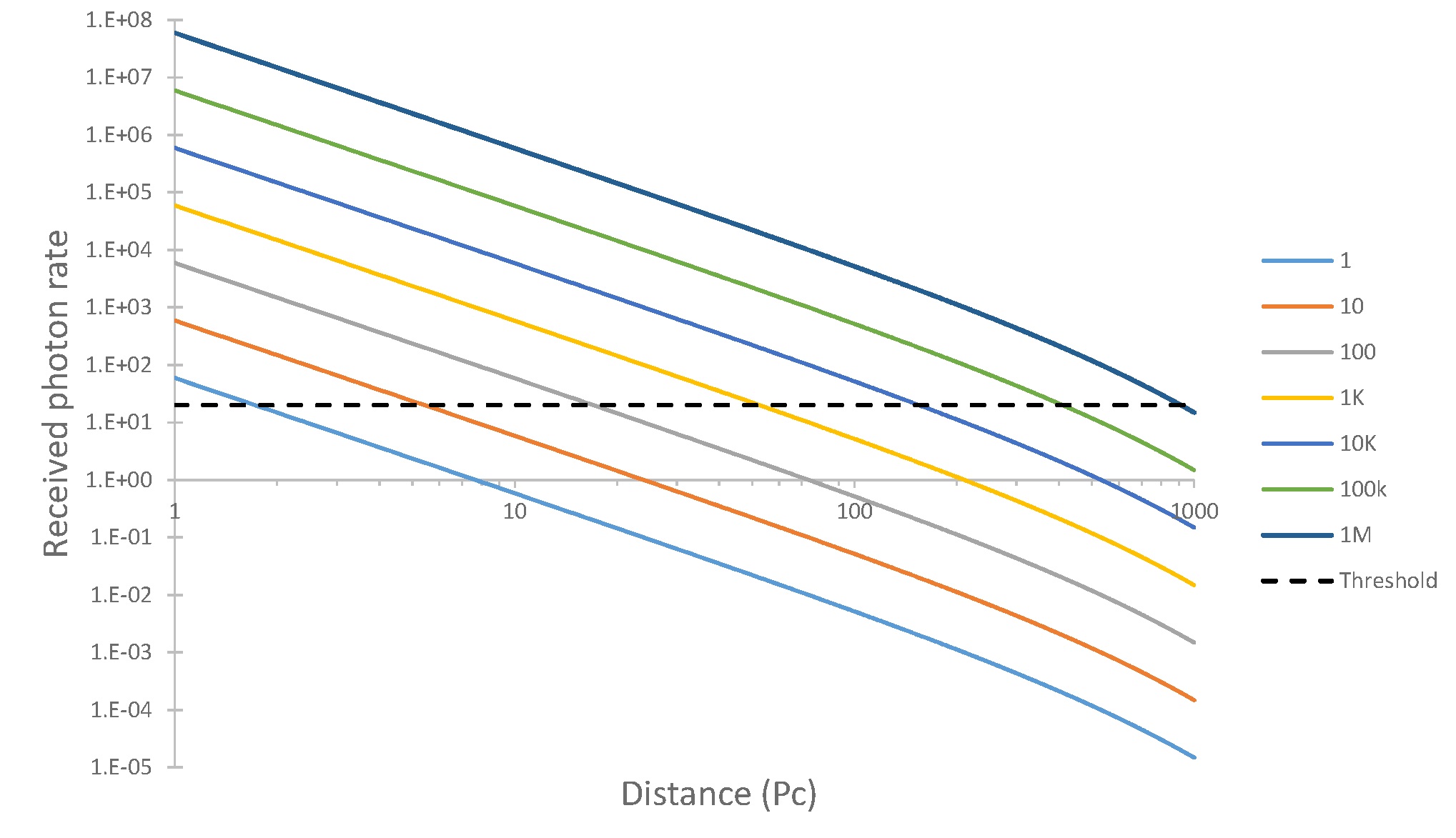}
    \caption{A plot of received photon number vs distance for a range of laser powers assuming a generic representative wavelength of 600nm (logarithmic scale). The horizontal line represents the threshold for detection. }
    \label{fig:Power}
\end{figure}
The transmit and receive aperture diameters are taken to be 10m. The received number of photons per second into a 10m diameter telescope is shown in the plot in figure\ref{fig:Power} for a laser wavelength of 600nm. This plot includes the effect of extinction by interstellar dust as given in \cite{Hippke2018}. Shorter wavelengths diffract less and increase the received photon rate but require more accurate pointing. A threshold rate for 5 sigma detection is drawn at 20 photons per second. It is interesting to note that a laser directed from our nearest neighbor the Alpha Centauri system  would only require a power of 1W to be detectable.  A 100kW laser should be detectable at a distance of 400Pc and a 1MW laser at 1kPc. 
It is worth noting that this may be an underestimate of the signal strength. The calculation uses an average photon number across the width of the beam, defined by the null annulus around the beam centre (this contains 84\% of the total beam power) but the peak at the centre can be twice as high as the average.  As pointed out by \cite{forgan2014can} is highly unlikely that we will intercept a beam by accident and therefore any detection will most likely arise from deliberate targeting. In such a scenario it is to be expected that beam pointing will be highly proficient. The use of a large synthetic aperture could also benefit the received photon number and reduce the energy cost to the sender.  
\subsection{Effectiveness}
The coherence detection technique is discussed in comparison to the dispersive spectroscopic techniques because it is the the only alternative method that can detect continuous laser sources. The values for expected sensitivity given in table \ref{tab:my_results} compare favourably with the spectroscopy measurements with sensitivities below 1 photon $s^{-1} m^{-2}$ but should be considered only as indicative of the potential of the technique. The spectroscopic method as described by \cite{Tellis2017} has the signal spread over multiple detecting pixels (a point spread function (PSF) width of 4.3 pixels) which is fundamentally limited by the system optics. With coherence detection the signal is spread across frequency and is limited by the resolution of the FFT - hence signal improves with increasing sample size. Comparison of the detected number of laser photons with the expected Poissonian noise level is shown in the Ratio column and approaches unity because the signal is spread over few frequency channels - it is effectively a narrower PSF.  Thus the threshold as calculated here could be improved by a longer observation time and also by a larger FFT size. 
Of course the clumping of counts into sampling periods is an artifice, made for convenient use of the FFT. The data exists as discrete events with a timing resolution determined by detector deadtime and more sophisticated analysis algorithms can be conceived to make use of this. The timing resolution also makes it possible to detect pulsed laser sources by looking for coincidences between the 2 detectors. This would work best if the pulse repetition rate is greater than the modulation frequency. The interferometer tends to direct coherent photons down the same arm so coincidences between arms are unlikely. This could be rectified by placing 2 detectors in each arm and looking for coincidences that occur in the correct arm relative to the interferometer setting. If however the detectors were photon number counting detectors this would not be an issue and a mean photon bunch can be extracted from information contained by the 2 detectors. This should also lead to a determination of the pulse repetition rate.
The data rate generated by such a system is given by the number of photons detected which is dependent upon the overall bandwidth and collection area. The critical factor is the spectral bandwidth per channel as this defines the noise level that must be exceeded, hence a modest spectrometer with a resolution of 0.1nm and an array of APDs ought to be both cost effective and very sensitive. The fifth row of table \ref{tab:my_results} shows a modelling outcome assuming a collection area of 1m$^{2}$. This has a higher sensitivity of 5.26 photons s$^{-1}$m$^{-2}$ but the ratio to the Poisson level is still good.  The Poisson nature of the statistics means that many such small systems could be added together to achieve an equivalent sensitivity to the larger telescope system.  The economy of scale gained from developing many smaller systems might suggest the search could move away from large telescopes and equip an array of smaller telescopes with a photon sensitive coherence detector.  
\subsection{Issues}
Up to this point we have assumed close to ideal circumstances of operation but at some point real world issues must be considered. Scintillation is an effect of the refractive index variations in the atmosphere which can lead a reduction in image quality and temporal variations of intensity \cite{andrews2005laser}. The so-called coherence length of the atmosphere is given by the Fried parameter \cite{fried1966optical}
\begin{equation}
    r_{0}=2.1[1.46 k^{2}L C_{n}^{2}]^{-3/5}
\end{equation}
where $k=2\pi /\lambda$, where $\lambda$ is the wavelength, L is the atmospheric path length, $C_{n}^{2}$ is the refractive index structure constant. In conditions of weak turbulence $C_{n}^{2}$ is of order $10^{-15}m^{-2/3}$ and a vertical view through the atmosphere is around 8km effective thickness. The Fried parameter is the effective scale at which an atmospheric patch produces a phase difference of 1 Radian. This distance is typically 10-20cm at observatory locations and smaller at ground level \cite{Lou2012}.  A full calculation of the effects of scintillation are beyond the scope of this paper and require further investigation. However it is appropriate to discuss the implications. The phase change upon a photon transiting the atmosphere into a large telescope will clearly be dependent upon entry time and position. This could have very negative implications, especially with the sparse photon rate that is expected as it could effectively scramble the phase relationship between consecutive photons. The effect of this atmospheric phase scrambling is what leads to degradation of image quality (reduction in the point spread function) and this has been overcome to some extent through the use of adaptive optics which can correct the spatial phase variance across the input. Thus there are 2 scenarios where coherence detection would work - in space-based telescopes where there is no atmosphere, and ground based telescopes utilising adaptive optics (it could be problematic using adaptive optics which makes use of a laser guide star!). There is a third possibility. The Fried parameter essentially represents a distance where the phase does not change polarity, thus if the aperture were of the same order as the fried parameter spatial phase variations would be negated somewhat. However this would represent a small collection aperture and many such apertures would be required to give a large collection area, but as we have already stated the addition of signals from multiple apertures is a feasible approach. This concept clearly requires more investigation.      
The second real world issue is one of cost. Discounting a space based telescope the components needed to construct a coherence detection system (APD array, spectrometer, electronics) are commercially available - including adaptive optics systems.  Thus either the adaptation of an existing telescope with an interferometer plus detector array, or the construction of an array of smaller telescope based modules are economically achievable aims and not requiring of an international scale development programme. 
Finally it has been suggested \footnote{by the referee of this paper} that such a system could be tested by making use of ground based laser ranging systems targeting satellites or the lunar based retro-reflectors. This would indeed provide an useful testbed for examining photon sensitive coherence detection. 

\section{Conclusion}
Attempts so far to detect technosignature optical signals originating from ETIs have focused on detecting enhanced spectral brightness using a high resolution spectrometer, or detecting short pulses using coincidence detection. In this paper it is proposed to look for coherence in an optical signal through the use of a modulating asymmetric MZI with a pair of photon sensitive detectors. This approach will detect continuous wave laser signals. Modelling suggests that for a limited  spectral bandwidth (0.1nm) and a 10m diameter collection telescope, the sensitivity would exceed the spectrometer approach where the spectral resolution is almost 2 orders of magnitude finer. The asymmetry of the MZI can be designed to ensure that atomic emission lines are not detected and mistaken for laser emission.   In addition the time-tagging of photon arrival times allows for coincidence detection which can identify pulsed sources as well as continuous sources - not possible by other means. The recording of each photon's arrival time allows a variety of algorithms to be applied to the data to extract signals. It also shows temporal variations in the  arrival times which will be useful in discriminating home grown signals. As well as incorporation into large area collections systems, a modular approach building a scalable system from many smaller collections areas is viable.

\bibliographystyle{unsrt}

\end{document}